\documentclass[twocolumn,prl,aps]{revtex4}
\usepackage{amsmath}
\usepackage{graphicx}
\usepackage{sistyle}

\begin{document}

\title{Interaction of Independent Single Photons based on Integrated Nonlinear Optics}

\author{T. Guerreiro$^1$, E. Pomarico$^1$, B. Sanguinetti$^1$, N. Sangouard$^1$, J. S. Pelc$^2$, C. Langrock$^2$, M. M. Fejer$^2$, H. Zbinden$^1$, R.~ T.~ Thew$^1$, N. Gisin$^1$}
\address{
$^1$Group of Applied Physics, University of Geneva, Switzerland
\\
$^2$E. L. Ginzton Laboratory, Stanford University, 348 Via Pueblo Mall, Stanford, California 94305, USA}

\begin{abstract}
Photons are ideal carriers of quantum information, as they can be easily created and can travel long distances without being affected by decoherence. For this reason, they are well suited for quantum communication \cite{Gisin2007}.
However, the interaction between single photons is negligible under most circumstances. Realising such an interaction is not only fundamentally fascinating but holds great potential for emerging technologies. It has recently been shown that even weak optical nonlinearities between single photons can be used to perform important quantum communication tasks more efficiently than methods based on linear optics \cite{Sangouard2011}, which have fundamental limitations~\cite{Kok2000}.
Nonlinear optical effects at single photon levels in atomic media have been studied~\cite{Schmidt1996,Chuang1995} and demonstrated~\cite{Turchette1995,Birnbaum2005,Pritchard2010,Peyronel2012} but these are neither flexible nor compatible with quantum communications as they impose restrictions on photons' wavelengths and bandwidths. Here we use a high efficiency nonlinear waveguide \cite{Parameswaran2002,Tanzilli2012} to observe the sum-frequency generation between a single photon and a single-photon level coherent state from two independent sources. The use of an integrated, room-temperature device and telecom wavelengths makes this approach to photon-photon interaction well adapted to long distance quantum communication, moving quantum nonlinear optics one step further towards complex quantum networks and future applications such as device independent quantum key distribution.
\end{abstract}

\maketitle

\begin{figure}[h!]
\centering
\includegraphics[width=0.8\columnwidth]{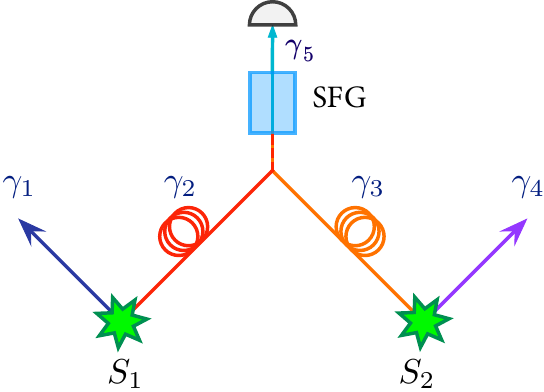}
\caption{Concept: Sum-frequency generation of photons from separate sources. A photon from each of two pair sources $S_1$ and $S_2$ is sent over a distance to a nonlinear waveguide. This waveguide performs the sum frequency generation (SFG) of photons $\gamma_2$ and $\gamma_3$ and outputs a photon $\gamma_5$. Detecting this photon heralds the distant presence of photons $\gamma_1$ and $\gamma_2$. Critically, the SFG process will work only for $\gamma_2$ and $\gamma_3$ and not for two photons coming from the same source. This feature would allow for faithful entanglement swapping, which is  otherwise impossible to perform with two probabilistic pair sources and linear optics.}
%Two-photon sum-frequency generation process: two input telecom photons b and c interact, mixing frequencies and producing a near infrared photon, e. One can imagine that photons b and c come from far apart, and their conversion heralds the entanglement of two other photons, a and d. In our experiment, one source produces pairs of single photons, while the other produces a single photon level coherent state.}
\label{photon_gate}
\end{figure}

The potential of parametric interactions used for quantum information processing has been demonstrated in a variety of interesting experiments~\cite{Kim2001, Dayan2005,Langford2011}. Although these interactions have been shown to preserve coherence~\cite{Giorgi2003, Tanzilli2005, Curtz2010}, they are generally performed using strong fields~\cite{Roussev2004,Vandevender2004,Thew2008}. It is only recently that parametric effects such as cross phase modulation~\cite{Matsuda2009,Lo2010} and spontaneous downconversion~\cite{Hubel2010, Shalm2013}  have been observed with a single photon level pump.
We take the next step and realise a photon-photon interaction, which can enable some fascinating experiments. For example, \figurename~\ref{photon_gate} shows how the sum-frequency generation (SFG) of two photons $\gamma_2$ and $\gamma_3$ from independent SPDC  sources can herald the presence (and entanglement) of two distant photons $\gamma_1$ and $\gamma_4$, as proposed in~\cite{Sangouard2011}.
The observation of a parametric effect between two single photons has been hindered by the inefficiency of the process in common nonlinear crystals.

%If the sources $S_1$ and $S_2$ produce entangled pairs, the detection of an SFG photon $\gamma_5$ can herald the entanglement between $\gamma_1$ and $\gamma_4$.
%Parametric processes, such as sum-frequency generation, are  generally performed with strong fields \cite{Roussev2004,Vandevender2004,Thew2008}, and have been shown to preserve coherence \cite{Giorgi2003, Tanzilli2005, Curtz2010}. The potential of  parametric interactions used for quantum information processing has been demonstrated in a variety of interesting experiments~\cite{Kim2001, Dayan2005,Langford2011}, but it is only recently that a parametric effect (spontaneous downconversion) has been demonstrated using a single-photon pump~\cite{Hubel2010, Shalm2013}. The nonlinear interaction between two single photons from independent  sources, as represented in~\figurename~\ref{photon_gate}, remains to be demonstrated.

% and used to perform a full Bell state measurement \cite{Kim2001}. In addition, they have been used to demonstrate interesting effects such as a nonlinear interaction with an ultrahigh flux of entangled photons~\cite{Dayan2005}, alternative approaches to frequency conversion \cite{Langford2011} and spontaneous downconversion pumped by a single photon enabling the generation of photon triplets~\cite{Hubel2010, Shalm2013}. However sum-frequency generation has never been demonstrated with two independent single photons.

\begin{figure*}[!ht]
\centering
\includegraphics[width=2\columnwidth]{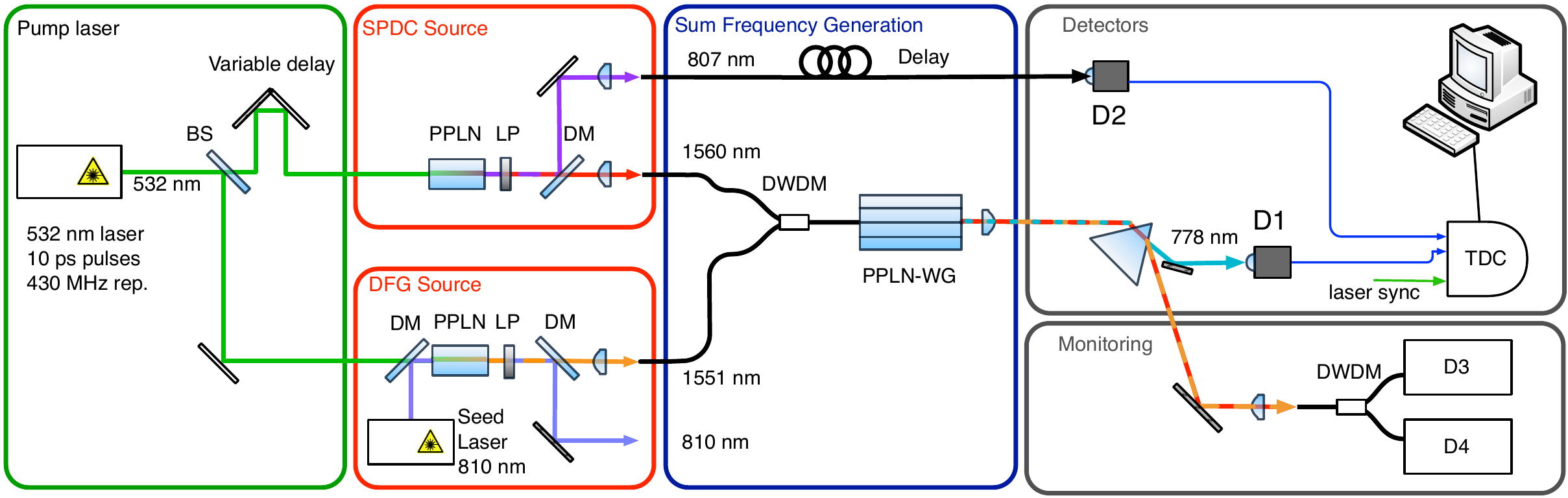}
\caption{Experimental setup. A mode locked laser that generates \SI{10}{ps} pulses at \SI{532}{nm} with a repetition rate of \SI{430}{MHz} is used to pump the two sources. The SPDC source consists of a Periodically Poled Lithium Niobate (PPLN) crystal set to produce pairs of photons at \SI{807}{nm} and \SI{1560}{nm}. These are deterministically separated by a dichroic mirror (DM) and collected into optical fibres. The remaining pump light is removed with a long pass filter (LP). The DFG source consists of a similar PPLN crystal seeded by a continuous wave laser at \SI{810}{nm} producing coherent state pulses at \SI{1551}{nm} by difference frequency generation. These pulses are combined into the same fibre via a \SI{200}{GHz} 3 port Dense Wavelength Divsion Multiplexer (DWDM) (ACPhotonics) and directed to the PPLN waveguide (WG). The PPLN-WG is \SI{4.5}{cm} long and has a second harmonic generation efficiency of \SI{41}{\%/W.cm^2} at \SI{1556}{nm}. The upconverted light is separated from the remaining telecom light via a prism, and sent to D1, a free space Si detector. The remaining unconverted telecom light is demultiplexed by another DWDM and sent to detectors D3 and D4, two free running InGaAs detectors. D3 and D4 are used for monitoring the coupling stability during the experiment and for optimising the transmission of the input fields through the waveguide by optimising their polarization. We record threefold coincidences between detectors D1, D2 and the laser clock signal using a time-to-digital converter (QuTools).}
\label{setup}
\end{figure*}

In our experiment we increase the interaction cross-section by strongly confining the photons, both spatially and temporally, over a long interaction length. The spatial confinement is achieved with a state-of-the-art nonlinear waveguide~\cite{Parameswaran2002,Tanzilli2012}, whilst the temporal confinement is obtained by using pulsed sources~\cite{Pomarico2012}. The efficiency of the process is proportional to the square of the waveguide length $L^2$, and inversely proportional to the duration of the input photons. $L$ is limited by the group velocity dispersion between the input photons and the unconverted photon~\cite{Sangouard2011}.
We maximise the SFG efficiency by matching the spectro-temporal characteristics of the single photons with the phase matching constraints of the waveguide. A \SI{4}{cm} waveguide and \SI{10}{ps} photons satisfy these conditions and are well suited to long distance quantum communication.

A schematic of the experimental setup is shown in \figurename~\ref{setup}. A \SI{532}{nm} mode locked laser produces pulses which pump two distinct sources. The first source produces pairs of photons by spontaneous parametric down conversion at \SI{807}{nm} and \SI{1560}{nm} (SPDC source). 
%This source has a probability of generating a pair per pulse of 0.1 in the mode defined by the optical fibres over the entire spectrum. 
Further details can be found in \cite{Pomarico2012}. The second source produces weak coherent state pulses at \SI{1551}{nm} by difference frequency generation (DFG source). The process is stimulated by a \SI{810}{nm} continuous wave seed laser. The average number of photons in the coherent state pulse can be adjusted by changing the seed and pump powers. All photons are coupled into single-mode fibres.

\begin{figure}[h!]
\centering
\includegraphics[scale=0.8]{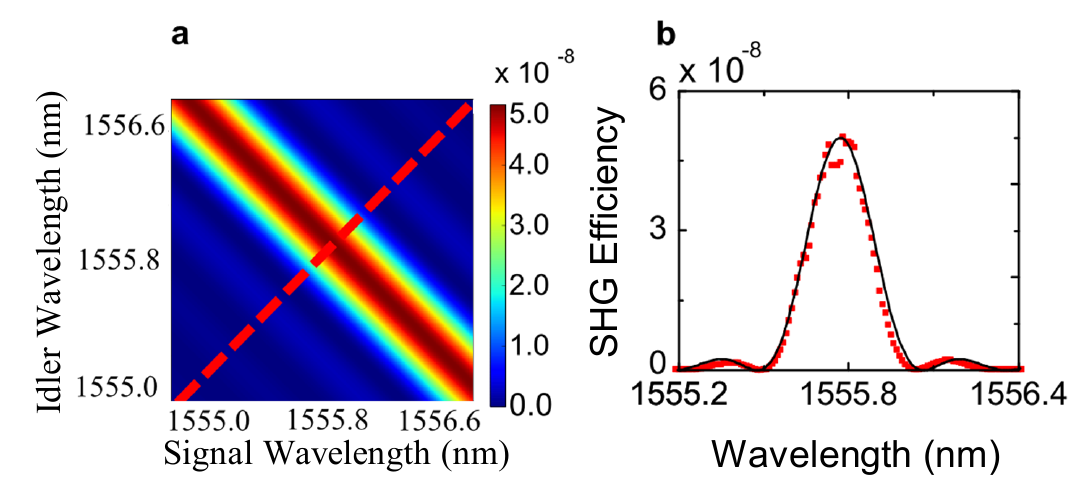}
\caption{\textbf{a}, Calculation of the waveguide efficiency taking into account the quasi-phase matching conditions. \textbf{b}, Measured values for the efficiency of second harmonic generation (red points) compared to the theoretical prediction (solid line), showing an almost ideal phase matching of the PPLN waveguide.}
\label{fig_scan}
\end{figure}

The telecom photons generated by the SPDC source are combined with the coherent state pulses from the DFG source using a  wavelength division multiplexer (DWDM). We verified the single photon nature of the SPDC source by measuring the conditional second order correlation function of the telecom photon after the DWDM to be $ g^{(2)}(0) = 0.03 $.

%In this measurement a free running Si detector triggers two gated InGaAs detectors positioned after a 50/50~beam splitter. From the single rates at the Si detector, InGaAs detectors and coincidence rates, we measured $ g^{(2)}(0) = 0.03 $, confirming the single photon nature of the input fields.

The photons are then sent to a fibre pigtailed reverse proton exchange Type $ 0 $~PPLN waveguide  \SI{4.5}{cm} long~\cite{Parameswaran2002}. This waveguide produces SFG of the input fields according to the phase matching conditions shown in  \figurename~\ref{fig_scan}. 
%The dashed line in the figure represents a section of the phase matching corresponding to second harmonic generation, where the two input beams have the same wavelength. 
The overall system efficiency for second harmonic generation (SHG) is measured  to be \SI{41}{\%/W.cm^2} at \SI{1556}{nm}, and used to estimate the SFG efficiency as described in the Supplementary Information. In addition to high efficiency, the waveguide exhibits almost ideal phase matching, as can be seen from figure \ref{fig_scan}\textbf{b}, as well as a high coupling of the fibre to the waveguide of $ 70 \% $.
 
\begin{figure}[h!]
\centering
\includegraphics[scale=1]{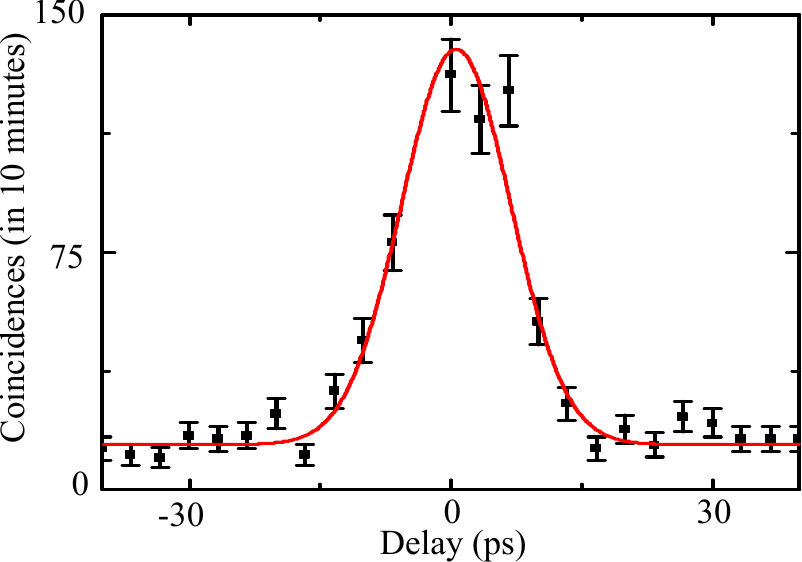}
\caption{Temporal overlap of input signals at the waveguide measured by introducing a delay between the photons' arrival times. The fit (solid line) gives a FWHM \SI{14.8}{ps}, which corresponds to a convolution of two pulses of \SI{10}{ps} from our laser. The spectra of the SPDC and of the DFG are, respectively, \SI{1.2}{nm} and \SI{0.8}{nm}, corresponding to  \SI{6.76}{ps} and \SI{10.03}{ps}. This indicates that to within the resolution of our measurement, our photons are close to being single mode.}
\label{fig_antidip}
\end{figure}

To verify the signature of our photon-photon interaction we record threefold coincidences between detectors D1 and D2 (both Si detectors), and the laser clock signal. When an upconverted photon is detected at D1 (\SI{3.5}{Hz} dark counts,  $ 62 \% $ detection efficiency at \SI{780}{nm}), an electric signal is sent to D2 (probability of dark count per gate $10^{-3}$, detection efficiency $ 40 \% $ at \SI{810}{nm}) \cite{Lunghi2012} opening a \SI{10}{ns} detection window. Conditioning the upconversion events on the laser clock signal helps to reduce the noise. %, and is realised with a home-made electronic card designed to deal with the 430~MHz rate of the laser.
%%%%%%%%%%%%%%%%%%%%%%%%%%%%%%%%%%%%%%%%%%%%%%%%%%%%%%%%%%%%%%%%%%%%%%%%%
We ensure that the photons arrive at the same time inside the waveguide by moving a motorised delay. \figurename~\ref{fig_antidip} shows the upconverted signal as a function of the delay between the photons. When performing this temporal alignment, the mean number of photons in the coherent state was increased to 25 per pulse. Each point of \figurename~\ref{fig_antidip} corresponds to the number of threefold coincidences between D1, D2 and the laser clock that occur over 10 minutes. The FHWM of the graph seen in FIG \ref{fig_antidip} is \SI{14.8}{ps}, which corresponds to the convolution of two \SI{10}{ps} pulses from the pump laser. From the spectra of the photons, which are \SI{1.2}{nm} for the SPDC and \SI{0.8}{nm} for the DFG we can deduce their coherence times, respectively \SI{6.76}{ps} and \SI{10.03}{ps}. This is a good indication that our photons are close to being pure. 

\begin{figure}[h!]
\includegraphics[scale=1]{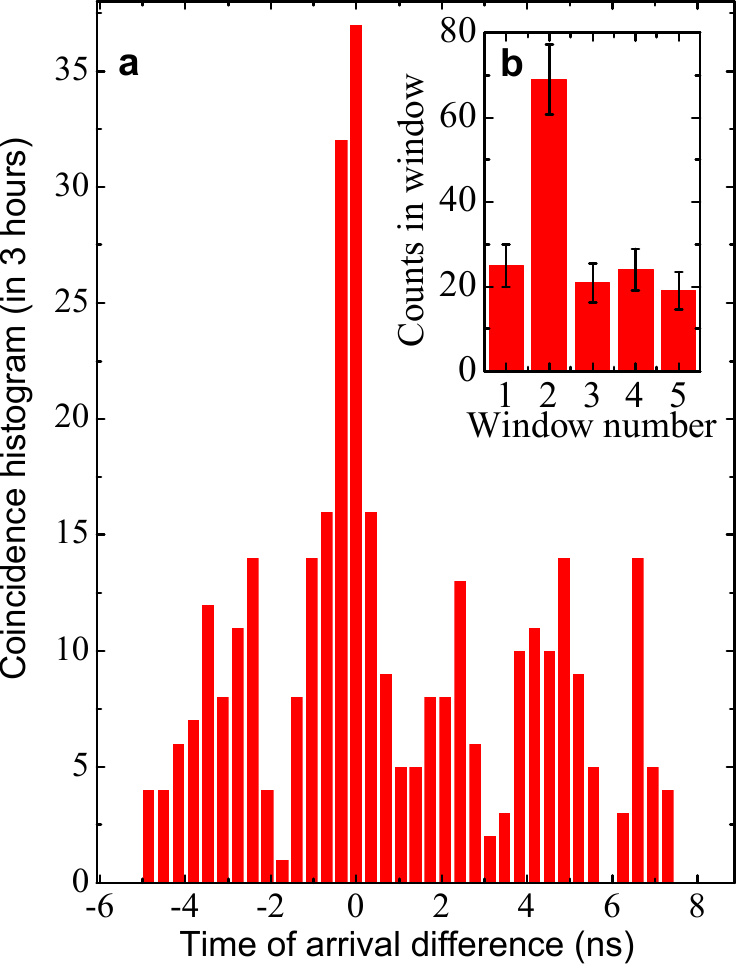}
\centering
\caption{\textbf{a}, Coincidence histogram between D1 and D2 showing strict time correlation between the heralding upconverted photon and the \SI{807}{nm} photon. The spread of the coincidence peak is mainly dominated by detector jitter. Side peaks are due to the periodicity of the laser. \textbf{b}, Bar chart histogram, obtained by integrating the two central bins of each peak.}
\label{fig_coin_laser}
\end{figure}

Once the SPDC and DFG sources have been characterised we  set the temporal delay to zero and measure the performance of the nonlinear interaction. For this measurement the coherent state had a mean number of 1.7~photons per pulse inside the waveguide. A histogram of arrival time differences is shown in FIG. \ref{fig_coin_laser}\textbf{a} (each bin corresponds to \SI{0.32}{ns}). The main peak is the signature of photon-photon conversion. It is also possible to see side peaks, which correspond to a dark count at D1 due to intrinsic noise of the detector followed by a detection of a photon at D2 (see the Supplementary Information). The periodicity of these side peaks corresponds to the period of the pump laser.

To more clearly see the signal to noise characteristics of the experiment we integrate over the events in the two central bins for each peak. This is shown in FIG. \ref{fig_coin_laser}\textbf{b}, where a peak with a signal to noise of 2 can be seen.

The coincidence rate between D1 and D2 was 25~$\pm$~5~counts per hour. To determine the efficiency of the SFG we can use this rate along with other independently measured parameters from our setup. We estimate the overall efficiency of the process at the single photon level to be $ \eta_{SFG} = (1.5 \pm 0.3) \times 10^{-8} $. Alternatively, using the measurement of second harmonic generation (SHG) efficiency, the calculation of the SFG efficiency shown in FIG. \ref{fig_scan} and accounting for the bandwidth of the interacting beams we estimated the efficiency to be $ 1.56 \times 10^{-8} $, which agrees well with the value estimated from the measured data. We highlight that this is the overall conversion efficiency, which includes the effects of coupling into the waveguide, internal losses and losses through the setup up to D1. Correcting for all of these losses we obtain the intrinsic device efficiency of $ (2.6 \pm 0.5) \times 10^{-8} $.

We have demonstrated the nonlinear interaction between a single photon and a single photon level coherent state. Such single-photon level parametric interactions open new perspectives for emerging quantum technologies. At the level of efficiency ($1.5 \times10^{-8} $) demonstrated here, the technique is competitive with linear optics protocols \cite{Sliwa2003, Barz2010, Wagenknecht2010}, and offers new possibilities such as heralding entanglement at a distance~\cite{Sangouard2011}. Also, unlike linear optics, there is significant scope for improvements as higher nonlinearities are realised. Work in this field is advancing rapidly: materials with higher nonlinear coefficients \cite{Kemlin2011} as well as methods for tighter field confinement~\cite{Kurimura2006}. The use of an integrated, solid state, room temperature device and a flexible choice of wavelengths will further aid the applicability of this type of system in future quantum communication technologies and beyond.

\textbf{Acknowledgments}: We are thankful to Anthony Martin for helpful discussions. This work was supported by the Swiss NCCR-QSIT and by the European project Q-ESSENCE.

\bibliography{SFG_ENDNOTE_EXPORT}

\section*{Supplementary Information}

\subsection*{Evaluation of the number of photons in the coherent~state}
To evaluate the number of photons per pulse in the coherent state, we measure the average power $ P_{\alpha} $ at the output of the DWDM. The average number of photons per pulse at this point is then given by

\begin{center}
$ \bar{n} =  \dfrac{\lambda P_{\alpha}}{hc} \times \dfrac{1}{f} $
\end{center}
where  $ f $ is the laser repetition rate of 430 MHz. To have the number of photons inside the waveguide, we multiply $ \bar{n} $ by the overall transmission of the setup from the DWDM to the interior of the waveguide, which includes the coupling of the pigtail inside the waveguide of $ 70 \% $. The overall transmission is $ 64 \% $.

\subsection*{Noise characterisation} 
\begin{figure}[h!]
\centering
\includegraphics[scale=0.8]{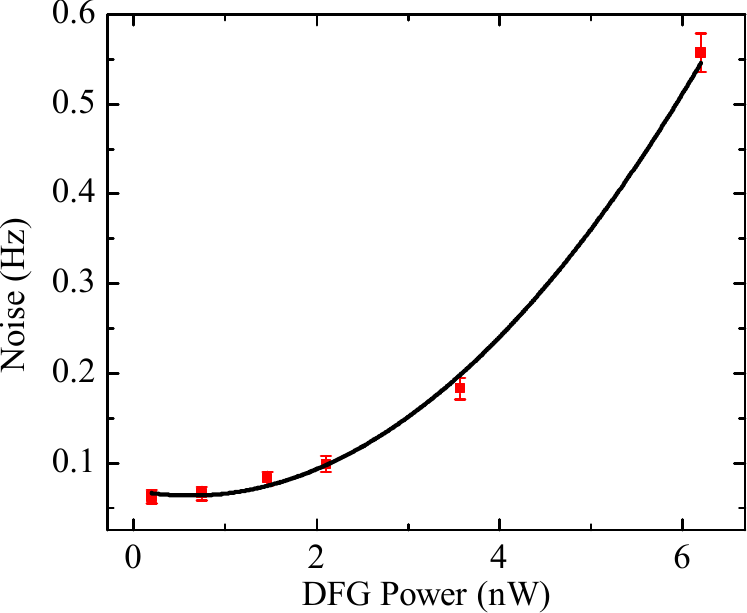}
\caption{Quadratic behaviour of noise in detector D2 as a function of DFG source power.}
\label{noise}
\end{figure}

Understanding the origin of the side peaks present in the graph of FIG. \ref{fig_coin_laser} is crucial. To do this we blocked the telecom photon coming from the SPDC source but not the coherent state from the DFG source, and recorded threefold coincidences between D1, D2 and the laser clock. The scaling of such noise in detector D2 as a function of the average power in the coherent state can be seen in FIG. \ref{noise}. Each point in the graph corresponds to a coincidence histogram integrated over 20~minutes. The quadratic behaviour of this noise suggests a possible contribution of second harmonic generation (SHG) from the \SI{1551}{nm} pulses to these side peaks.  

\begin{figure}[h!]
\centering
\includegraphics[scale=0.8]{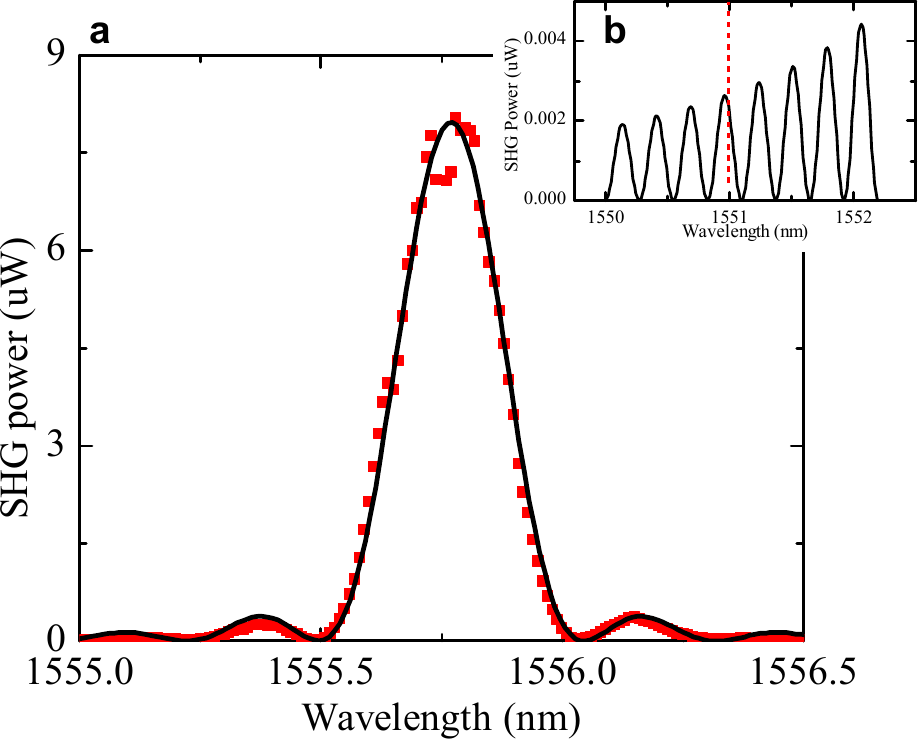}
\caption{ \textbf{a}, Second harmonic generation spectrum. Solid line: fit of the data. \textbf{b}, Zoom of fit, showing the predicted intensity of SHG at 1551~nm.}
\label{SHG_fit}
\end{figure}

To evaluate this, we estimate the effective SHG efficiency from the second harmonic spectrum seen in FIG. \ref{SHG_fit}. The peak value of such spectrum corresponds to a measured efficiency of \SI{41}{\%/W.cm^2}. From a fit of such a spectrum, we conclude that the effective SHG efficiency for 1551 nm is $ \eta_{SHG}(1551)~=$~2.35~$ 10^{-4} \times$~\SI{41}{\%/W.cm^2}.

By taking into account this effective efficiency we can estimate the expected rates at detector D1 due to SHG of the coherent state pulses. These rates are simply

\begin{center}
$ R_{SHG}(P_{\alpha}) = (P_{\alpha} \times \mu)^{2} \times \eta_{SHG}(1551) \times L^{2} \times \dfrac{\lambda}{hc} $
\end{center}
where $ \mu $ is the coupling efficiency of the coherent state into the optical fibre, which was measured to be $ 76 \% $ and $ L $ is the length of the waveguide.  The result of this estimation, compared with the actual measured values can be seen in table \ref{results}.

\begin{table}[h!]
\begin{center}
\begin{tabular}{l*{6}{c}r}
$ P_{\alpha} $ (nW)    & Measured (Hz) & Calculated (Hz)  \\
\hline
6.20 & 15.30 & 18.60   \\
3.54 & 5.70 & 6.60   \\
2.10 & 3.80 & 2.30   \\
1.47 & 4.00 & 1.10   \\
0.75 & 3.20 & 0.30   \\
0.00 & 3.50 & 0.00   \\
\end{tabular}
\caption{\label{results} Measured and calculated counts at D1 as a function of the power of the coherent state. As the power is reduced, the SHG contribution to clicks at D1 falls below the detector noise level. During the experiment we operate in this regime.}
\end{center}
\end{table}
From such an analysis we can conclude that the second harmonic generation  contribution to the noise  at the single photon level can be neglected. The side peaks are then dominated by coincidence between a dark count at D1 and a detection at D2. This confirms that a detection of an upconverted photon does not come from conversion of two photons from the same DFG pulse.

\subsection*{SFG efficiency measurement}
Using the data shown in FIG. \ref{fig_coin_laser} we can extract the rate of coincidences between D1 and D2, $ R_{SFG} $. By combining these with other numbers from the setup, independently characterised, we can then extract the overall sum-frequency generation efficiency from a single photon measurement. The numbers used to obtain this efficiency can be seen in table \ref{parameters}.

\begin{table}[h!]
\begin{center}
\begin{tabular}{l*{6}{c}r}
\multicolumn{2}{c}{\textbf{SPDC source}}                                                   \\
\hline
\hline
$ p_{SPDC} $ & 0.03 ph/mode \\
$ t_{807} $ & 0.4    \\
$ t_{grating} $ & 0.7    \\
$ \eta_{807} $ & 0.4    \\
$ t^{SPDC}_{DWDM} $ & 0.86   \\
$ t_{bandwidth} $ & 0.4 \\
\\
\multicolumn{2}{c}{\textbf{DFG source}}                                                    \\
\hline
\hline
$ p_{DFG} $ & 1.7 ph/mode \\
$ t^{DFG}_{DWDM} $ & 0.96   \\
$ \eta_{780} $ & 0.6 \\
\end{tabular}
\caption{\label{parameters} Key parameters for the setup which were independently characterised, where $ p $ stands for number of photons per mode, $ t $ for transmission. Each particular index refers to a different optical element.}
\end{center}
\end{table}

The sum-frequency generation efficiency is then
\begin{center}
$ \hat{\eta}_{SFG} = \dfrac{R_{SFG}}{\beta} $
\end{center} 
where $ \beta $ is the product of all the quantities in Table~\ref{parameters} times the laser repetition rate of 430~MHz. From the experimental data we obtained $ R_{SFG} = 25 \pm 5 $ counts per hour, yielding an overall efficiency of $ \eta_{SFG}~=~(1.5~\pm~0.3)~\times ~10^{-8} $.

\subsection*{SFG efficiency estimation}
It is natural to ask whether the value found for the SFG efficiency agrees with the value for the efficiency of the SHG, measured classically, shown in FIG. \ref{fig_scan}. To do that, we modelled the phase matching conditions of the waveguide using the appropriate Sellmeier equations \cite{Jundt1997}.

The peak value of the SHG efficiency, corresponding to a wavelength of \SI{1556}{nm}, is \SI{41}{\%/W.cm^2}. Given an efficiency measured at the classical level $ \eta $ we can obtain the corresponding value at the single photon level $ \hat{\eta} $ using the equation \cite{Sangouard2011}

\begin{center}
$ \hat{\eta}(\lambda) = \dfrac{\eta(\lambda)}{2} \times \dfrac{hc}{\lambda} \times \dfrac{\Delta\hat{\nu}L}{tbp} $
\end{center}
where for our system $L = $ \SI{4.5}{cm}, $ \Delta\hat{\nu} = $ \SI{296}{GHz.cm} and $tbp=~0.66$. To give an example, the peak level for the SHG efficiency then reads $ \hat{\eta}(1556) = 5 \times 10^{-8} $ which is of the same order of magnitude of the SFG efficiency obtained experimentally. This estimation, however, did not take into account the bandwidth of the interacting fields. 

To take this into consideration, we use the matrix shown in FIG. \ref{fig_scan} to obtain $ \hat{\eta}(\lambda_{s},\lambda_{i}) $, the efficiency as a function of the wavelengths of the input fields. We then integrate over the spectra of the interacting beams, normalised to the area, denoted by $ p_{s}(\lambda_{s}) $ and $ p_{i}(\lambda_{i}) $. The total effective efficiency reads

\begin{equation*}
\hat{\eta}^{eff}_{SFG} = \iint p_{s}(\lambda_{s}) p_{i}(\lambda_{i}) \hat{\eta}(\lambda_{s},\lambda_{i}) d\lambda_{s} d\lambda_{i} = 1.56 \times 10^{-8}
\end{equation*}
which is in agreement with the value found from the measured data.

\end{document}